\documentclass[review,3p,fleqn,times]{elsarticle}
\journal{Current Opinion in Systems Biology}

\usepackage{amsmath}
\usepackage{graphicx}
\usepackage{geometry}
\usepackage{natbib}

\usepackage[utf8]{inputenc}
\usepackage{hyperref}

\usepackage{soul}
\usepackage{color}

\usepackage{caption}
\usepackage{placeins}
\usepackage{enumitem}

\begin{document}

\begin{frontmatter}
\title{On structural and practical identifiability}
\author[IP,FDM,CIBSS]{Franz-Georg Wieland}
\author[IP,FDM]{Adrian L.\ Hauber}
\author[IP,FDM]{Marcus Rosenblatt}
\author[IP,FDM,CIBSS]{Christian Tönsing}
\author[IP,FDM,CIBSS]{Jens Timmer\corref{mycorrespondingauthor}}
\cortext[mycorrespondingauthor]{Corresponding author}
\ead{jeti@fdm.uni-freiburg.de}
\address[IP]{Institute of Physics, University of Freiburg, Hermann-Herder-Str.\ 3, 79104 Freiburg, Germany}
\address[FDM]{Freiburg Center for Data Analysis and Modelling (FDM), University of Freiburg, Ernst-Zermelo-Str.\ 1, 79104 Freiburg, Germany}
\address[CIBSS]{Centre for Integrative Biological Signalling Studies (CIBSS), University of Freiburg, Schänzlestr.\ 18, 79104 Freiburg, Germany}

\begin{abstract}
We discuss issues of structural and practical identifiability of partially observed differential equations which are often applied in systems biology. The development of mathematical methods to investigate structural non-identifiability has a long tradition. Computationally efficient methods to detect and cure it have been developed recently. Practical non-identifiability on the other hand has not been investigated at the same conceptually clear level. We argue that practical identifiability is more challenging than structural identifiability when it comes to modelling experimental data. We discuss that the classical approach based on the Fisher information matrix has severe shortcomings. As an alternative, we propose using the profile likelihood, which is a powerful approach to detect and resolve practical non-identifiability.
\vspace{0.7cm}\\
\textbf{Highlights}
\begin{itemize}[noitemsep]
\item With recent advances, identifying structural identifiability is no longer a major issue
\item Practical identifiability is still challenging
\item Fisher information matrix is misleading
\item Profile likelihood can solve practical identifiability challenge
\end{itemize}
\end{abstract}



\end{frontmatter}

\begin{figure*}[ht!]
\begin{center}
\textbf{Graphical abstract}
\vspace{1.7cm}
 \includegraphics[width=\textwidth]{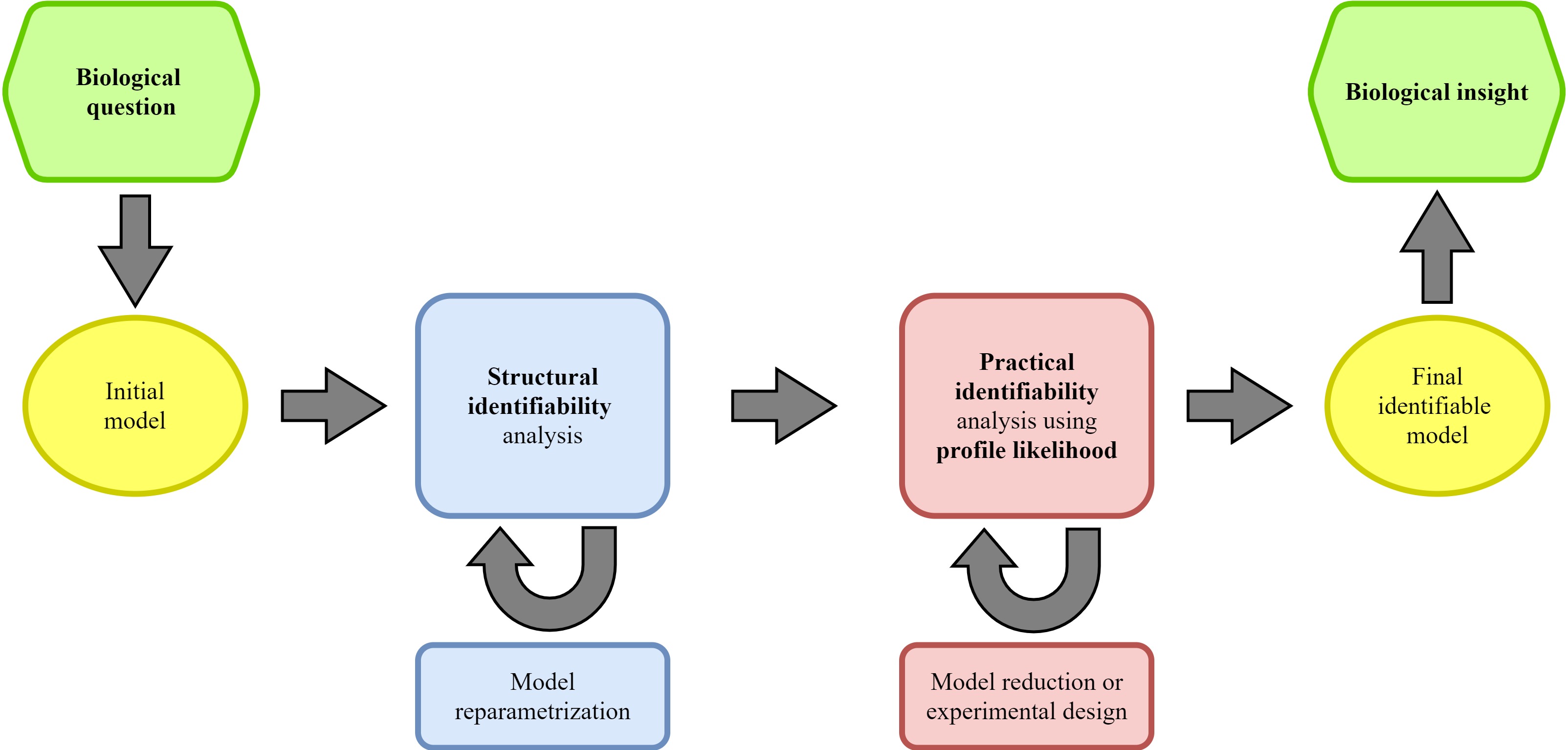}
 \end{center}
\end{figure*}
\FloatBarrier

\section*{Introduction}
\subsection*{Biological modelling and Box's statement}
Traditional biological reasoning often is rather qualitative, descriptive, and static, which results e.g.\ in cell biology in so-called ``pathway cartoons''. Mathematical models based on differential equations can help to turn these into a quantitative, predictive, and dynamic understanding of the underlying system. Discussing modelling in general, in 1979, George E.P. Box coined his famous statement: ``All models are wrong, but some are useful'' \cite{box79}. While the former part of the quote is intuitively clear, since every model necessarily poses a simplification of reality, the latter highlights the importance of assessing what constitutes a \emph{useful} model.

\subsection*{Bad, good and useful models}
Three properties comprise a \emph{useful} model. Firstly, it has to capture the main effects of the question of interest, i.e.\ describe the data with reasonable accuracy, and neglect the rest. Secondly, a \emph{useful} model has to make experimentally falsifiable predictions in order to be testable. Models that exhibit these two properties are \emph{good} models. Thirdly, the model should enable to gather insights about the biological system.
In a typical modelling process, one starts off with an initial model based on current biological knowledge. Usually, this model cannot explain the data and therefore is a \emph{bad} model. Based on biological intuition and trial-and-error, one increases the model complexity until the data can be fitted. Often, this leads to an over-parameterised model that over-fits the data. The parameters of such a model and in turn its predictions are not well-determined and it thus remains a \emph{bad} model. 

The path from such a \emph{bad} model towards a \emph{good} model is laborious: additional data needs to be measured and integrated, the model complexity needs to be reduced and balanced to the available data, or a combination of both.
This process needs to be iterated until a \emph{good} model is found, which has well determined parameters and predictions.

However, such a \emph{good} model also needs to deliver biological insights in order to be \emph{useful}. Only this third property turns a \emph{good} model into a \emph{useful} model.
In this sense, the final goal of mathematical modelling in systems biology is not the model itself but to use the model to understand biology. One example of how a model can be used to gain biological insight, which would be unattainable by merely assessing the data by itself, was given by \citet{becker08}.

\subsection*{Parameter identifiability}
The concept of identifiability is strongly linked to the transition from \emph{bad} models to \emph{good} models. Identifiability analysis is necessary to create \emph{good} models that can describe the data and have well-determined parameters and predictions. It is especially important when modelling biological systems because the limited amount and quality of the experimental data with large measurement noise in only partially observed systems often leads to \emph{bad} models during the modelling process.
Concerning identifiability, one distinguishes between structural identifiability dealing with inherently indeterminable parameters due to the model structure itself, and practical identifiability, dealing with insufficiently informative measurements to determine the parameters with adequate precision.

\section*{Partially observed dynamical systems}
A biological system is translated into ordinary differential equations (ODEs) 
\begin{equation} 
\dot{x} = f(x,p,u), \label{eq:ode}
\end{equation}
comprising $n$ model states $x(t)$, unknown parameters $p$ to be estimated from time-resolved experimental data, and external stimuli $u(t)$. Since data is often recorded on a relative scale, scaling and offset parameters for background corrections need to be estimated in parallel. Furthermore, in typical applications not all components of a cell-biological system can be measured, e.g.\ because of the limited availability or restricted capability of antibodies to discriminate between un-phosphorylated, i.e.\ inactive, and phosphorylated, i.e.\ active, proteins. Thus, an observation function $g(\cdot)$ is required that maps the internal states $x$ to the observations: 
\begin{equation} 
y=g(x,p,t). \label{eq:observables}
\end{equation}
Typically, the dimension $m$ of $y$ is smaller than the dimension $n$ of $x$. We are therefore dealing with parameter estimation in partially observed systems. Moreover, in systems biology, these ODE models are typically stiff, non-linear, sparse and non-autonomous, and the discrete time observations are noisy.

Parameter estimation is usually performed based on the weighted residual sum of squares, the negative log-likelihood assuming Gaussian errors 
\begin{equation}
 \chi^2_\text{res} (p) = \sum^{m}_{k=1} \sum^{d_k}_{l=1} \left( \frac{y^D_{kl} - g_k(p,t_l) }{\sigma^D_{kl}}  \right)^2,
 \label{Eq:Chi2Equation}
\end{equation}
to determine the agreement of experimental data with the model trajectories, where $y^D_{kl}$ and $\sigma^D_{kl}$ represent $d_k$ data points and measurement errors at time points $t_l$ for each observable.
A common point estimate for the best parameter vector is
the maximum likelihood estimator
\begin{equation}
\label{Eq:MLE}
 \hat{p} = \arg \min \left[ \chi^2_\text{res}(p) \right].
\end{equation}

\section*{Structural identifiability}
\subsection*{Definition of structural identifiability and connection to observability}

Partially observed dynamical systems often exhibit structural non-identifiability. A model is structurally identifiable, if a unique parameterisation exists for any given model output. A parameter $p_i$ is globally structurally identifiable \cite{Walter1997}, if for all parameter vectors $p$, it holds
\begin{equation}
 y(p) = y(p') \hspace{0.2cm} \Rightarrow \hspace{0.2cm} p_i = p'_i \,.
\end{equation}
An individual parameter $p_i$ is structurally non-identifiable, if changing the parameter does not necessarily alter the model trajectory $y$, because the changes can be fully compensated by altering other parameters. Local structural identifiability of a parameter is defined by reducing the definition to a neighbourhood~$v(p)$ instead of the entire parameter space. A model is structurally identifiable, if all of its parameters are structurally identifiable. Multiple related definitions for structural identifiability exist, for a comprehensive discussion see a recent overview \cite{Anstett-Collin2020}.

A structurally non-identifiable parameter implies the existence of a manifold in parameter space upon which the trajectory $y$ is unchanged. However, on this manifold the dynamic variables $x$ of the model can change, e.g.\ by a scaling factor, and are thus not uniquely determinable. This is denoted as non-observability, a concept closely related to parameter non-identifiability \cite{Kalman1959,Bellman1970,chappell90,Villaverde2019,Schmitt2020}. 

\subsection*{A priori analysis of structural identifiability}
Two basic approaches exist to assess structural identifiability of non-linear dynamic models. \emph{A priori} methods only use the model definition, while \emph{a posteriori} methods use the available data to find non-identifiable parameters. Many \emph{a priori} algorithms have been developed based on a variety of approaches. Powerful methods use Lie group theory, since non-identifiabilities are closely related to symmetries in the system \cite{Yates2009,merkt14,Villaverde2019b,massonis20}.  Furthermore, a variety of notable methods exist, which are based on power series expansion \cite{pohjanpalo78}, generating series \cite{Walter1982,Ligon2018}, semi-numerical approaches \cite{sedoglavic02,Karlsson2012}, differential algebra \cite{Ljung1994,Saccomani2003,Bellu2007,Meshkat2014,Thomaseth2018,Varghese2018,Saccomani2019,Hong2019, Hong2020}, differential geometry \cite{Villaverde2019a}, and numerical algebraic geometry \cite{Bates2019}. For reviews of some of these approaches, see \cite{Villaverde2019a,Chis2011,raue14a}. Many of these approaches, especially the early developed methods, can only be applied to rather low-dimensional systems because of their computational complexity. Thus, recent developments have mainly focused on improving the computational efficiency of the algorithms, e.g.\ by local sensitivity calculations. 

As a promising example, \citet{joubert20} proposed a comprehensive and computationally fast pipeline to cure structural non-identifiabilities by re-parameterisation of the model in a five-step procedure: (i) a numerical identifiability analysis based on sensitivities, (ii) symbolic identifiability calculations for the low-dimensional candidates from (i), this renders the procedure fast, (iii) computation of new model parameters, this step is not unique, but requires decisions of the modeler, (iv) simplify the original model leading to a lower dimensional parameter vector, and finally (v) check the identifiability of the re-parameterised model. In an application to a model with 21 states and 75 parameters, two groups of non-identifiable parameters were detected and the model was re-parameterised within minutes.

\subsection*{Analysis of structural identifiability using experimental data}
In contrast to the aforementioned methods, \emph{a posteriori} methods use the available data to perform identifiability analysis. They infer structural non-identifiability based on model fits to experimental data. Similar to some sensitivity-based \emph{a priori} approaches, these approaches only assess local structural identifiability.

One approach by \citet{hengl07a} suggested to perform numerous fits and investigate non-parametrically whether the final parameter estimates form a low-dimensional manifold in parameter space. This approach also allows to disentangle different sets of coupled non-identifiable parameters. 

\begin{figure*}[ht!]
  \centering
  \includegraphics[width=\textwidth]{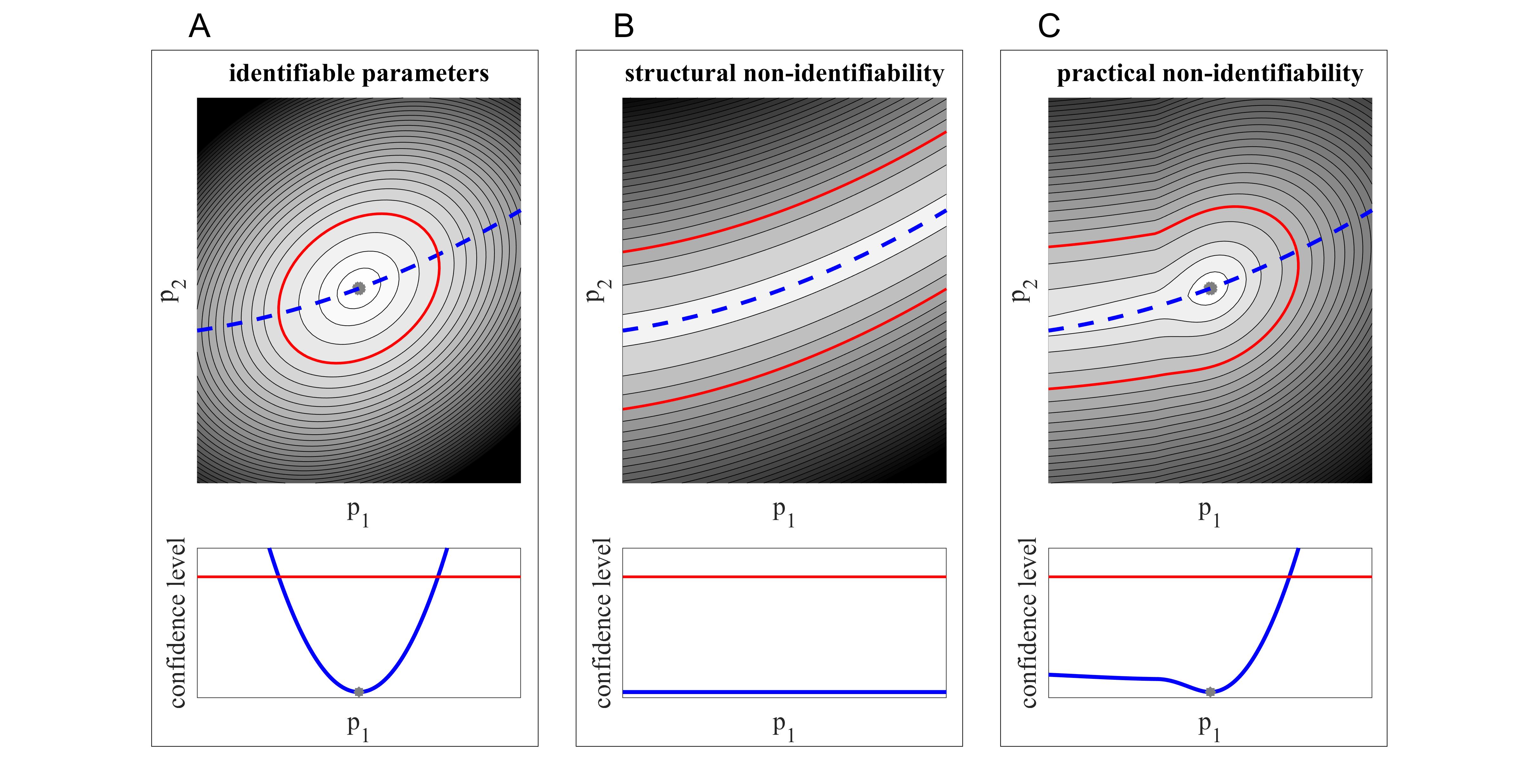}
  \captionof{figure}{\textbf{Illustrative example of likelihood contour plots and profile likelihood for an identifiable parameter and structurally and practically non-identifiable parameters.} Subfigures (A), (B), and (C) show contour plots of $\chi^2_\text{res}$ above as well as the profile likelihood versus the parameter below. Lighter colours in the contour plots signify a lower value of $\chi^2_\text{res}$. Thresholds for confidence intervals corresponding to a confidence level of 95 \% are shown in red and plotted both in the contour plots and the profile likelihood plots. The lowest value of $\chi^2_\text{res}$ is denoted by a grey asterisk in both the contour plot and the profile likelihood plot.
  For the identifiable parameter (A) the profile likelihood reaches both an upper and lower threshold thus leading to a finite confidence interval.
  For the structurally non-identifiable parameter (B) the profile likelihood is completely flat, thus yielding infinite confidence intervals. In the contour plot this translates to a flat path, along which $\chi^2_\text{res}$ does not change. 
  The practically non-identifiable parameter (C) shows an infinite extension of the low $\chi^2_\text{res}$ region for lower values of the parameter, never reaching the 95 \% confidence interval threshold. In contrast, a finite upper confidence bound can be derived.}
  \label{Fig:ProfilesAndContour}
\end{figure*}

An informative and successful method is based on the profile likelihood \cite{Murphy2000}. The idea of the profile likelihood is to vary one parameter $p_i$ after the other around the maximum likelihood estimate (Equation (\ref{Eq:MLE})) and re-optimise the remaining ones
\begin{equation}
 \label{PL}
     \mathrm{PL} \left( p_i \right) = \min_{p_{j\neq i}} \left[ \chi^2_\text{res} \left( p \right) \right]\,.
\end{equation}
For the two-parameter examples in Figure~\ref{Fig:ProfilesAndContour}, the blue dashed lines show the path in the parameter space determined by Equation (\ref{PL}). Figure~\ref{Fig:ProfilesAndContour}A shows the profile likelihood of an identifiable parameter. For a structurally non-identifiable parameter the profile likelihood yields a flat line as shown in Figure~\ref{Fig:ProfilesAndContour}B. Plotting the remaining parameters along the profiled parameter reveals which parameters are coupled to the non-identifiable one \cite{raue09a}. The profile likelihood was recently extended to include two-dimensional profiles to allow for the identification of parameter interdependence \cite{Brastein2019}.

Profile likelihood calculation can be computationally demanding for larger systems due to the numerical re-optimisation. Addressing this issue, a fast \emph{a posteriori} method to test identifiability without the need to calculate complete profiles using radial penalisation was recently developed \cite{Kreutz2018}.

\FloatBarrier
Structural non-identifiability can also be investigated \emph{a posteriori} by a Bayesian Markov chain Monte Carlo (MCMC) sampling approach. However, for non-identifiable systems efficient mixing and thus convergence of the Markov chains is difficult \cite{raue13a}. This problem can be cured by informative priors but these would mask the problem and should only be implemented if they are based on actual biological insights and prior information. One recent application in the field identified a minimal subset of reactions in a signalling network with a combination of parallel tempering and LASSO regression methods \cite{Gupta2020}.

\subsection*{Re-parameterising structurally non-identifiable models}
Given the recent advances in the computational efficiency of methods, we essentially consider determining structural identifiability no longer a bottleneck in the modelling of non-linear dynamic systems with ODEs. For models with a high number of connected structurally non-identifiable parameters, finding and resolving these structural non-identifiabilities can still be challenging. This is often the case if the number of observed states is much lower than the number of dynamic states. When the structurally non-identifiable parameters are determined, the problem is usually fixed by a re-parameterisation of the model. In the simplest case this is accomplished by fixing some of the involved parameters to a certain value. The price to be paid is typically that the information about the scale of some components is lost which can limit the predictive power of the model. Nevertheless, biologically meaningful re-parameterisation of the models after finding non-identifiabilites remains a challenging task (G. Massonis et al.\,, arXiv:2012.09826v2).

\section*{Practical identifiability}
\subsection*{From structural to practical identifiability}
Structural identifiability implies practical identifiability only for an infinite amount of data with zero noise. Practical identifiability is important for obtaining precise parameter estimates. Moreover, it is especially crucial to ensure that model predictions are well-determined. It is analyzed increasingly often to judge a model's predictivity \cite{Busch2015,David2019,Duchesne2019,Zhou2020,Johnson2020}. The notion of practical identifiability has been rather vague in the literature, mainly referring to \emph{large confidence intervals} \cite{Nihtila1977,Holmberg1982,Miao2011}. Some approaches exist that define practical identifiability as a combination of model structure and experimental protocol without actual data \cite{Gontier2020, Saccomani2018}. In contrast, we consider a combination of model and data as practically identifiable if the confidence intervals of all estimated parameters are of finite size \cite{raue09a}.

\subsection*{Parameter confidence intervals and identifiability}
The profile likelihood (Equation~(\ref{PL})) provides a proper assessment of confidence intervals of estimated parameters in ODE models (Figure \ref{Fig:ProfilesAndContour}) by
\begin{equation}
\mathit{CI}_{\mathrm{PL}} \left( p_i \right) = \bigl\{ p_i \:|\: \mathrm{PL} \left( p_i \right)  \leq \chi^2_\text{res} \left( \hat{p} \right) + \Delta_\alpha  \bigr\},
\label{eq:PLconfInt}
\end{equation}

where $\Delta_\alpha$ denotes the $\alpha$ quantile of the $\chi^2$ distribution with $\mathit{df}=1$ degrees of freedom for point-wise confidence intervals \cite{Murphy2000}.

\begin{figure*}[ht!]
  \centering
  \includegraphics[width=\textwidth]{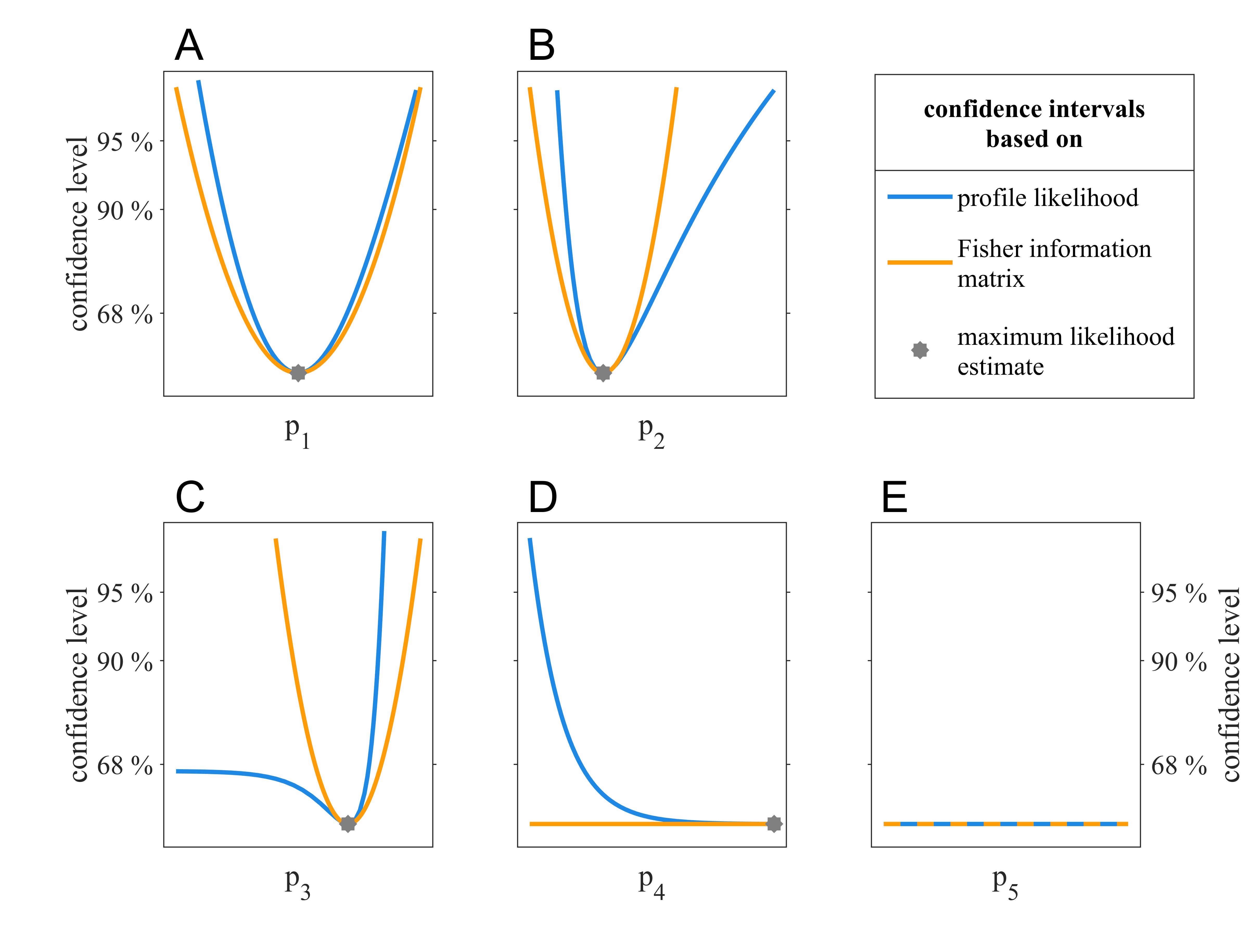}
  \captionof{figure}{\textbf{Parameter confidence intervals based on Fisher information matrix and profile likelihood.} Confidence intervals for five parameters based on profile likelihood (blue) and on quadratic approximation using the Fisher information matrix (FIM) (orange). FIM-based confidence intervals have two major problems. Firstly, due to the non-linearity of the underlying systems, the Cramér-Rao bound on the error is invalid and thus the FIM-based confidence intervals become uncontrollable for a finite amount of measurements. While in (A) the FIM-based interval is larger than the profile likelihood-based interval, in (B) it is smaller. Secondly, FIM based intervals are insensitive to practical non-identifiabilities. In (C), the FIM-based confidence interval is finite and thus the practically non-identifiable parameter is not detected. In (D), the practically non-identifiable parameter leads to a flat FIM-based interval, wrongly suggesting structural non-identifiability. While the structurally non-identifiable parameter in (E) is correctly detected, similarly to (D) the calculation of the FIM is challenging due to its singularity in flat likelihood landscapes. The parameters (A)-(D) are adapted from two applications in synthetic biology ((A),(B),(C) from \citet{Schneider2021}, (D) from \citet{Ochoa-Fernandez2020}). Parameter (E) is from a minimal non-identifiable toy model. Grey asterisks signify the maximum likelihood estimate of the parameter.}
  \label{Fig:ConfidenceIntervalsFIMvsPL}
\end{figure*}

The traditional method for determining confidence intervals based on the Fisher information matrix (FIM) leads to accurate confidence intervals for linear regression models. Since the solutions of all nontrivial ODE models are non-linear in their parameters, using this method for analysing identifiability of such models is questionable \cite{Neale1997}. Furthermore, in contrast to FIM-based confidence intervals, profile likelihood-based confidence intervals can be asymmetric and are invariant under re-parameterisations of the model, e.g.\ the often applied logarithmic transformation of the parameters. Figure \ref{Fig:ConfidenceIntervalsFIMvsPL} shows five parameters with FIM- and profile likelihood-based confidence intervals, mainly taken from applications in synthetic biology \cite{Ochoa-Fernandez2020,Schneider2021}.

Identifiability is obtained if all estimated parameters are structurally and practically identifiable, i.e.\ have finite confidence intervals. A non-identifiable parameter is called practically non-identifiable if the confidence interval becomes finite for a given confidence level by adding additional measurements for the existing observables (Figure \ref{Fig:ProfilesAndContour}C). By adding enough data, a practically non-identifiable parameter can be made identifiable.

\FloatBarrier

\subsection*{Bayesian methods for identifiability analysis}
Bayesian sampling approaches, e.g.~MCMC, can be used to assess practical identifiability \cite{Siekmann2012,Hines2014,Zuo2019}. This, however is only feasible if the model is structurally identifiable, since structural non-identifiabilities will lead to bad mixing of the sampling algorithms. Given a structurally identifiable model, MCMC sampling yields similar results as the profile likelihood analysis \cite{raue13a}. However, a recent application in a spatio-temporal reaction–diffusion model showed that it is one order of magnitude slower than the profile likelihood \cite{Simpson2020}. To the best of our knowledge, a comprehensive benchmark study comparing the two methods is so far missing.

\section*{Experimental design and model reduction}
\subsection*{Model predictions}

 To test the predictive power of a model, confidence intervals for the predictions can be computed. For this purpose, forward evaluations of the model are utilized, e.g.~bootstrap approaches \cite{Joshi2006} or sensitivity analysis \cite{Sachs1984}. They typically require large numerical efforts in the context of non-linear biological models with a high-dimensional parameter space. A more powerful approach is the \textit{prediction profile likelihood}
\begin{equation}
 \label{Eq:PPL}
     \mathrm{PPL} \left( z \right) = \min_{ p \in \{ p \:|\: g_\text{pred} \left(  p \right) = z \} } \hspace{0.05cm} \left[  \chi^2_\text{res} \left( p \right) \right] \,,
\end{equation}
which is obtained by minimising $\chi^2_\text{res}(p)$ (Equation (\ref{Eq:Chi2Equation})) under the constraint that the model response $g_\text{pred} \left(  p \right)$ is equal to the prediction $z$. The \textit{prediction profile likelihood} propagates the uncertainty from the experimental data to the prediction by exploring the prediction space instead of the parameter space \cite{kreutz12a}.

Model predictions have to be sufficiently precise to produce insights. For special cases, this can be achieved without identifiability \cite{Cedersund2016,Rateitschak2012}. If the model predictions are not of sufficient precision, one has two principal options to tailor the model complexity to the information content of the data: (i) measure additional data, corresponding to an increase of the dimension of the observation function $g$ in Equation~(\ref{eq:observables}), or (ii) reduce the model complexity according to the available data, corresponding to a decrease of the dimension of the parameter space and/or of the ODE system $f$ in Equation~(\ref{eq:ode}). Both options increase the practical identifiability of the model.

\subsection*{Achieving practical identifiability by new measurements with optimal experimental design}

Practical identifiability can be achieved by adding new data \cite{Johnson2020, Bhonsale2020}. The process of  determining the most informative targets and time points for the new measurements is known as \textit{optimal experimental design} and is frequently applied in different modelling fields, e.g.\ metabolic models \cite{Froysa2020}, animal science \cite{Munoz-Tamayo2018}, linear perturbation networks \cite{Gross2020} or synthetic biology \cite{Bandiera2020}. The task is related to the search of an additional measurement that contains the maximal information about the system or parts of it. For improving the identifiability of a specific parameter, the model trajectories along the corresponding parameter profile can be investigated \cite{raue10, Steiert2012}. Thereby, measurement points with maximal information content for the parameter of interest can be determined, which corresponds to trajectories with high spread. Similarly, the prediction profile likelihood (Equation (\ref{Eq:PPL})) determines the prediction uncertainty of the model at a potential new measurement time point \cite{kreutz12a} thus promoting the identifiability of the whole model. Measurement points with high prediction uncertainty are effective to constrain the model further, whereas measurements with a low prediction uncertainty are better suited for model selection purposes.

\subsection*{Achieving practical identifiability by reducing model complexity}

If measuring additional data is not feasible, the complexity of the model has to be reduced. One way is to fix parameter values or ratios of parameters by means of prior knowledge \cite{Tivay2020}, sensitivity analysis \cite{He2020, Zi2011} or profile likelihood \cite{Pironet2019}. However, fixing parameters can decrease the interpretative relevance of the model's predictions. 

Taking this into account, a systematic model reduction strategy that tailors model complexity to the available data was suggested by \citet{maiwald16a}. Based on likelihood profiles, they discuss four basic scenarios that are discriminated based on the profile likelihood by the combinations of: either (i) the profile flattens out for a logarithmised parameter going to infinity or (ii) to minus infinity, and either other parameters are (a) coupled to the investigated one or (b) not. For all four possible combinations, there is a cure. For case (i/a), one differential equation is replaced by an algebraic equation, for (i/b), states can be lumped, for (ii/a), a variable is fixed, leading to a structural non-identifiability that can be cured by the methods discussed above, and for (ii/b), a reaction can be removed from the model. This model reduction strategy has been applied e.g.\ in \cite{Ochoa-Fernandez2020,Schneider2021,toensing17a}. Independent of the applied method, model reduction steps, and in particular the conclusions thereof, should always be documented together with the model according to good scientific practice to facilitate reproducibility.

\section*{Conclusions}
Given the multitude of recently developed methods \cite{massonis20, Ligon2018, Hong2020, joubert20}, we consider the file of identifying structurally non-identifiable parameters as closed. Future research in this field could focus on identifying biologically plausible re-parameterisations of the model, for which no comprehensive method yet exists to our knowledge. Furthermore, the extension of the concept of identifiability to different model types, e.g.\ mixed effects models \cite{CucurullSanchez2019,Janzen2019}, is of interest.

\begin{figure*}[ht!]
  \centering
    \includegraphics[width=\textwidth]{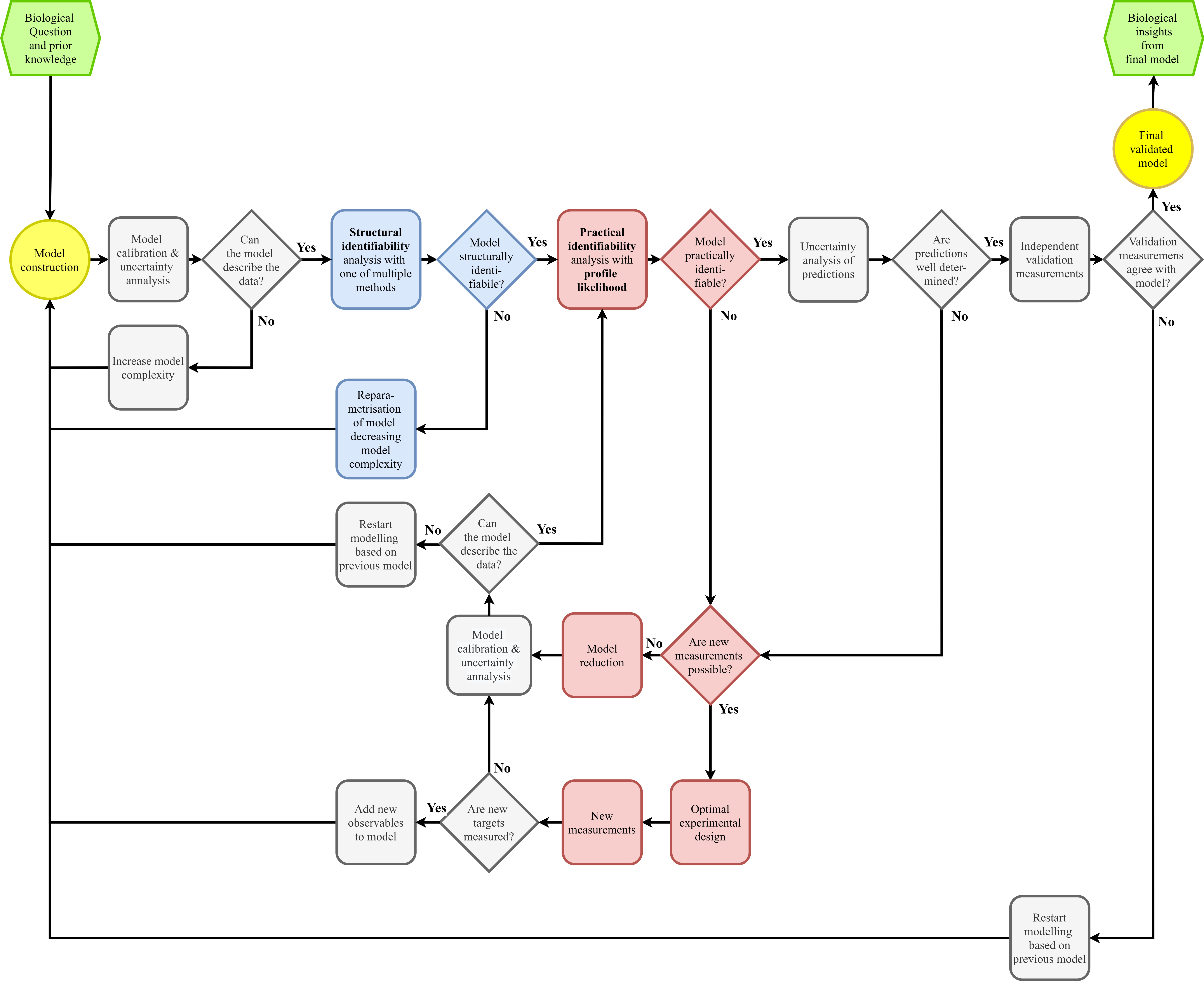}
  \title{Flowchart of the entire modelling process}
  \captionof{figure}{\textbf{Flowchart of the entire modelling process from initial to final model including identifiability analysis.} The modelling process begins with the inception of an initial model based on prior knowledge and the underlying biological research question. It ends with the final validated model and the biological insights it provides. The flowchart shows, how identifiability analysis is embedded into the overall modelling workflow. The topics discussed in this review related to structural identifiability (blue) and practical identifiability (red) are highlighted with colours in the flowchart. The remaining tiles in gray represent aspects that are beyond the scope of this review. The intricacy of the flowchart shows, that the path to biological insights requires multiple iterations of different methods. Identifiability analysis is an integral part of this workflow and should be performed to gain insights from predictive models with well-determined parameters. Furthermore, methods dealing with structural and practical identifiability should always be focused on ultimately progressing along the path towards biological insights.}
   \label{Fig:Flowchart of modelling process}
\end{figure*}

Achieving practical identifiability for model and data is more laborious in practice. Practical non-identifiabilities can be detected reliably, e.g.\ by the profile likelihood method \cite{raue14a}. In order to achieve identifiability, the model complexity has to be reduced or additional data must be added. Profile likelihood-based model reduction \cite{maiwald16a} and optimal experimental design \cite{Steiert2012} provide valuable methods for these purposes. A flowchart locating structural and practical identifiability analysis as discussed in this review within the entire modelling process is given in Figure \ref{Fig:Flowchart of modelling process}. 

Although the availability of advanced methods for the detection and cure of structural and practical non-identifiabilities is promising, two related challenges remain. In many applications identifiability analysis is not performed with state-of-the-art methods. Particularly, identifiability analysis based on the Fisher information matrix can be misleading in typical applications in systems biology. We propose a more consequent use of the discussed methods for structural identifiability and especially profile likelihood for practical identifiability analysis in order to check the limitations and predictive power of mathematical models. In summary, we believe the focal point of research in systems biology should always remain on the biological insights that can be gained from mathematical models which are structurally and practically identifiable.
\FloatBarrier

\section*{Conflict of interest statement}
The authors declare that they have no known competing financial interests or personal relationships that could have appeared to influence the work reported in this paper.
\section*{Acknowledgements}
This work was supported by the German Research Foundation (DFG) under Germany's Excellence Strategy (CIBSS – EXC-2189 – Project ID 390939984), the German Research Foundation (DFG) through grant 272983813/TRR 179, the Deutsche Krebshilfe (grant 70112355), the German Federal Ministry for Education and Research within the research network Systems Medicine of the Liver (LiSyM; grant 031L0048), and by the state of Baden-Württemberg through bwHPC and the German Research Foundation (DFG) through grant INST 35/1134-1 FUGG. 

The authors thank Hans Stigter for insightful feedback on the manuscript and Daniel Lill for his help in the writing process.

The graphical abstract and Figure \ref{Fig:Flowchart of modelling process} were created with draw.io.

\bibliography{BibliographyTimmerIdentifiability.bib}
\bibliographystyle{elsarticle-num-names-annote} 

\end{document}